\documentstyle[prl,aps,twocolumn]{revtex}
\tightenlines
\begin{document}
\twocolumn[\hsize\textwidth\columnwidth\hsize\csname
@twocolumnfalse\endcsname
\draft
\title{Proof of unconditional security of six-state
quantum key distribution scheme}
\author{Hoi-Kwong Lo}
\address{MagiQ Technologies Inc.,
275 Seventh Avenue,
26th Floor,
New York, NY 10001}

\date{\today}
%\preprint{quant-ph:}
\maketitle
\begin{abstract}

We prove the unconditional security of the standard six-state scheme for
quantum key distribution (QKD). We demonstrate its unconditional
security up to a bit error rate of
12.7 percents, by allowing only one-way classical
communications in the error correction/privacy amplification procedure
between Alice and Bob.
This shows a clear advantage of the six-state scheme over another
standard scheme---BB84,
which has been proven to be secure up to only about 11 percents, if only
one-way classical communications are allowed. 
Our proof technique is a generalization of that of Shor-Preskill's proof
of security of BB84. We show that a
advantage of the six-state scheme lies in the
Alice and Bob's ability to establish rigorously from their test
sample the non-trivial mutual information between the bit-flip
and phase error patterns. A modified version of the degenerate quantum codes
studied by DiVincenzo, Shor and Smolin is employed in
our proof.

\end{abstract}
\pacs{PACS Numbers:}
]
\narrowtext
\section{BB84}
\label{Intro}

Whereas conventional cryptography is often based on
some unproven computational assumptions, the security of quantum
key distribution  \cite{bb84,ekert,pt,gisin} \footnote{Quantum
cryptography, but not quantum key distribution per se,
was invented by Stephen Wiesner around 1970 in a paper that
remained unpublished until 1983 \cite{wiesner}.
Quantum key distribution is experimentally the most advanced
subfield of quantum information processing. Photons have been
transmitted over about 50km of commercial Telecom fibers
\cite{50km}  and over 1km of
open air \cite{air}.} is guaranteed
by the fundamental laws (particularly the uncertainty principle) of
quantum mechanics. The best-known
quantum cryptographic application is quantum key distribution (QKD)
whose goal is to allow two persons, Alice and Bob, to
communicate in perfect security in the presence of an eavesdropper, Eve.

A number of years had passed before rigorous and
convincing proofs
of security against the most general attack finally appeared.
Mayers \cite{mayersqkd} and
subsequently others \cite{others} have proven the security of the
standard Bennett and Brassard's BB84 scheme \cite{bb84}, a scheme that is
closer to a realistic experimental
situation. Unfortunately, those proofs are rather complex.
A proof by Lo and Chau \cite{qkd} has the advantage of being conceptually
simple, but it requires a quantum computer to implement.
Their proof built on earlier work on quantum privacy
amplification \cite{deutsch} and has subsequently been further
simplified \cite{simple}.
Recently, Shor and Preskill \cite{shorpre} have proposed a simple proof of
security of BB84 by combining and generalizing the insights in
Lo and Chau's \cite{qkd} and Mayers' \cite{mayersqkd} proofs.
Their proof also extends the
tolerable error rate of BB84 from about $7$ percents set by
Mayers' proof to about $11$ percents.\footnote{Mayers' proof also
permits the same extension.}

Other QKD schemes have also been proposed.\footnote{For instance,
an efficient four state scheme has been proposed and its
unconditional security was proven in \cite{eff}. Besides,
the security of a continuous variable (squeezed state)
QKD scheme has been proven by
Gottesman and Preskill \cite{squeeze} using
the same approach of Shor and Preskill's proof.}
A notable example is the six-state scheme proposed
by Bruss \cite{bruss}. Recently, a proof of security of
the six-state scheme has been proposed by Inamori \cite{inamori6},
which, unlike Shor and Preskill's proof of security of BB84,
requires two-way classical communications between Alice and Bob.

Until now, it was not obvious how to generalize
the proof technique of Shor-Preskill's proof, which
requires only one-way classical communications, to the six-state scheme.
The main goal of this paper is to provide precisely such a
simple proof of security for the six-state using
Shor and Preskill's approach. Our result shows
that, using only one-way classical communication,
six-state QKD scheme can be made secure up to an error rate of
about 12.7 percents. This is higher than the value of about 11 percents
in the case of Shor-Preskill's result in BB84, thus demonstrating
the advantage of the six-state scheme over BB84.
Our proof also
clarifies the symmetry structure employed in Shor and Preskill's proof.

A key idea of Shor-Preskill's proof is reduction:
Instead of tackling the security of BB84 directly,
they took an indirect path. They constructed a QKD scheme that employs
entanglement purification (i.e., it requires
a quantum computer to implement) and showed that such a
scheme is secure. Then, they showed that the security of
such an entanglement-purification-based QKD scheme implies the
security of BB84. In their proof, the bit-flip and phase errors
of the underlying entanglement purification protocol
may be totally uncorrelated. Therefore, in the worst
case situation, the bit-flip error syndromes tell the two
users nothing about the phase errors.

In this paper, we will follow
Shor-Preskill's approach for the case of a six-state QKD scheme.
We see a clear advantage of the six-state scheme over BB84:
As will be discussed in subsequent sections,
for the six-state scheme, one can show that in
the corresponding underlying entanglement-purification-based QKD
scheme, the bit-flip and phase errors are
{\it correlated}.\footnote{As shown in subsequent sections,
the six-state QKD scheme can be
made symmetric with respect to all three bases, $X$, $Y$
and $Z$. In the language of entanglement purification, this
corresponds to a so-called depolarizing channel. Therefore,
the bit-flip and phase errors are, indeed, correlated.}
In other words, the bit-flip error syndromes can be used to
reduce the conditional entropy of the phase error pattern.
This reduction in conditional entropy makes
the task of entanglement purification easier and
allows us to establish
the security of the six-state scheme up to an error rate of 12.7 percents.

This paper is organized as follows. In Section~2, we
review Shor-Preskill's proof. In Section~3, we study
the differences between BB84 and the six-state scheme,
emphasizing the ability by Alice and Bob to establish the
correlations between the bit-flip and phase error patterns
in the six-state scheme, but {\it not} in BB84.
In Section~4, our protocol for secure six-state QKD scheme is
given. Section~5 contains various concluding remarks.

\section{Shor-Preskill's proof}
In this section, we shall recapitulate briefly
Shor and Preskill's proof \cite{shorpre} of
security of BB84. A nice review of Shor and Preskill's proof
can be found in the early sections of
\cite{squeeze}.
Readers who are familar with
the subject can skip this section.
Before we go to the specifics, we should first review
the three major ingredients of their proof:
entanglement purification, classicalization
(i.e., quantum to classical reduction)
and CSS codes.

\subsection{Entanglement purification}
%Notice that, in the proof of security of QKD,
%we originally have a noisy quantum problem to
%consider. Entanglement purification removes
%noises from the problem and reduces
%a noisy quantum problem to a noiseless
%quantum problem.
Entanglement purification was first
studied by Bennett, DiVincenzo, Smolin and Wootters\cite{BDSW}
and its usage in QKD was first proposed
by Deutsch {\it et al.} \cite{deutsch}.
Suppose Alice prepares $n$ EPR pairs and sends the half of
each pair to Bob through a channel controlled by Eve.
Because of Eve's interference, the $n$ EPR pairs are
now noisy. However, Alice and Bob can purify from the $n$ imperfect pairs
a smaller number, say $m$, perfect EPR pairs, {\it provided that}
the channel is not too noisy.

\subsection{Classicalization}
\label{ss:classical}
A key question remains: how can one {\it verify} that the
channel is, indeed, not too noisy? This is not entirely trivial
because noise pattern of the channel is controlled
by Eve and does not have to be independent. Moreover,
the Einstein-Podolsky-Rosen paradox tells us that
it would be too naive to apply classical arguments blindly
to a quantum problem. This is where the classicalization
(quantum to classical reduction) idea of Lo and Chau \cite{qkd}
comes in.

The key idea is ``commuting observables'',
i.e., one should focus on observables that commute with each
other. For those observables, it is consistent to assign
probabilities to their {\it simultaneous} eigenstates and study
those probabilities by classical probability theory, particularly
classical random sampling theory. This leads to substantial
simplification of the original quantum problem.
[This ``commuting observables'' idea is the essence of
the stabilizer formalism of Gottesman \cite{stabilizer} and
Calderbank {\it et al.} \cite{calderbank}.]

More concretely, Alice and Bob can figure out the
error rate of the two (rectilinear or diagonal) bases
by random sampling. That is to say that, for each basis,
Alice and Bob select a random subset of test EPR pairs
and compare their polarizations of the two halves of a pair to
see if they agree. Mathematically, this is equivalent to
measuring the either operator $XX$ and $ZZ$,
where $X$ and $Z$ are respectively the Pauli matrices,
$\sigma_x$ and $\sigma_z$. The key observation here is
that $XX$ commutes with $ZZ$. Therefore, the commuting
observables idea indeed applies and probabilities to
the simultaneous eigenstates can be assigned to Alice and Bob's
state. 

With the above two ingredients---entanglement purification and
classicalization, one can prove the security of QKD by
intuitive classical argument \cite{qkd}.
Nonetheless, the resulting
protocols still require quantum computers to implement.
This is because a general entanglement purification protocol
requires a quantum computer for its implementation. It is
the following insight of Shor and Preskill \cite{shorpre}
that allows one to
implement a secure QKD scheme without a quantum computer.

\subsection{CSS codes}
Their proof makes essential use of the
Calderbank-Shor-Steane (CSS) code.
The CSS code has the
useful property that the error correction procedure for
the phase error is decoupled from that for the bit-flip
error. Clearly, bit-clip error correction is important to
ensure that Alice and Bob do share a common key.
However, Shor and Preskill made the following important
observation:
Since phase errors will not change the
bit value of their final key anyway, Alice and Bob have the
liberty of dropping the whole phase error correction
procedure altogether.
This is the fundamental reason why they can implement a CSS code-based
QKD scheme without a quantum computer.
General quantum error correcting codes can also be
used for QKD, but it is unclear how to implement those
QKD schemes without a quantum computer.

Even though the phase error correction procedure is
dropped in BB84, it is, nonetheless,
important that the phase error is, in principle, correctable
by the underlying quantum error correcting code because
only then can security be guaranteed by the quantum no-cloning
theorem. In other words, Alice and Bob do not need to
perform phase error correction. The very fact that Alice
and Bob {\it could} perform phase error
correction (if they had quantum computers) would be
enough to guarantee security of QKD.
The phase error correction procedure reduces the eavesdropper's
information on the key to an exponentially small amount in terms of
some security parameters. In other words, the phase error
correction is used for privacy amplification, whereas
the bit-flip error correction is used for error correction.
The remenant of the phase error correction procedure is a
``coset extraction'' procedure.
This point has been emphasized in \cite{shorpre} and will
be recapitulated below.

\subsection{Notation}
Having introduced the above
three major ingredients, we shall give
more specifics of the Shor-Preskill's proof.
We shall mostly use the notations in \cite{shorpre}.
For each qubit, we use a canonical basis, $ | 0 \rangle$
and $ | 1 \rangle$. Define also the basis, $ | + \rangle$
and $ | - \rangle$, where
$ | + \rangle = {  1\over \sqrt{2}} ( | 0 \rangle + | 1 \rangle )$
and $ | -\rangle = {  1\over \sqrt{2}} ( | 0 \rangle - | 1 \rangle )$.
The Hadamard transform, $H$, is a single qubit unitary transformation of
the form:

\begin{equation}
H = { 1 \over \sqrt{2} } \left( \begin{array}{cc} 1 & 1 \\
 1 & -1 \end{array} \right).
\label{e:hadamard}
\end{equation}
in the canoncial basis. It interchanges the bases $ | 0 \rangle$,
$ | 1 \rangle$ and $ | + \rangle$,
$ | - \rangle$.

Let us also
introduce Pauli matrices,
\begin{equation}
\begin{array}{ccc}
\sigma_x =
\left(
\begin{array}{cc} 0 & 1 \\1& 0 \end{array}
\right),
&
\sigma_y =
\left(
\begin{array}{cc} 0 & -i \\
 i & 0 \end{array}
\right),
&
\sigma_z =
\left(
\begin{array}{cc} 1 & 0 \\ 0 & -1\end{array}
\right).
\end{array}
\end{equation}
In what follows, we may simplify our notation and
denote the three Pauli matrices simply by
$X$, $Y$ and $Z$.

The Bell basis is an orthogonal basis for the quantum
state of two qubits. It has basis vectors,
\begin{eqnarray}
\Psi^{\pm}& =& { 1\over \sqrt{2}} ( | 0 1 \rangle \pm | 10 \rangle ) , 
\label{e:psi}\\
\Phi^{\pm}& =& { 1\over \sqrt{2}} ( | 0 0 \rangle \pm | 11 \rangle ).
\label{e:phi}
\end{eqnarray} 

\subsubsection{CSS codes}

Let us consider two classical binary codes, $C_1$ and $C_2$, such that,
\begin{equation}
\{0\} \subset C_2 \subset C_1 \subset F^n_2,
\end{equation}
where $F^n_2$ is the binary vector space of the $n$ bits and that
both $C_1$ and $C_2^{\perp}$, the dual of $C_2$ can correct
up to $t$ errors. A basis for the CSS code can be found as follows.
For each $v \in C_1$, define the vector
\begin{equation}
v \to { 1 \over  | C_2|^{1/2}} \sum_{w \in C_2} 
| v + w \rangle .
\label{eq:v1}
\end{equation}
Notice that $v_1 $ and $v_2$ give the same vector whenever
$v_1 - v_2 \in C_2$. In other words, the codeword of the CSS
code corresponds to the coset of $C_2$ in $C_1$.
Let $H_1$ be the parity check matrix for
the code $C_1$ and $H_2$ for $C_2^{\perp}$.

\subsection{Secure QKD based on entanglement purification}
Let us recapitulate the key point of Shor-Preskill's proof.
As a starting point of their paper, they
\cite{shorpre} proved the security of the following QKD scheme:

{\bf Protocol 1} (in \cite{shorpre}): {\bf Modified Lo-Chau}

(0) Alice and Bob decide on a large positive integer $n$,
a CSS code and a maximal number $e_{max}$ of check bit errors that they
tolerate in the protocol.

(1) Alice prepares $2n$ EPR pairs in the state $ (\Phi^+)^{2n}$.

(2) Alice picks a random $2n$-bit string $b$ and applies a
Hadamard transform $H$ on the second half of each EPR pair
for which (the component of) $b$ is $1$.

(3) Alice sends the second halves of the EPR pairs to Bob.

(4) Bob receives the qubits and publicly acknowledges the completion of
his reception.

(5) Alice selects randomly $n$ of the $2n$ EPR pairs to serve as check
bits to test for the eavesdropper, Eve,'s eavesdropping.

(6) Alice announces the bit string $b$ and which $n$ EPR
pairs are to be used as check bits.

(7) Bob performs a Hadamard transform on the qubits where (the component
of) $b$ is $1$.

(8) Alice and Bob each measure their halves of the $n$ check
EPR pairs in the $ | 0\rangle$, $| 1 \rangle$ basis and
broadcast their results. If more than $e_{max}$
check bits disagree, they abort. Otherwise, they proceed to
the next step.

(9) Alice and Bob each measure $\sigma_z^{[r]}$ for each row
$r \in H_1$ and  $\sigma_x^{[r]}$ for each row
$r \in H_2$. They broadcast their results. Bob transforms
his state accordingly to obtain $m$ nearly perfect EPR pairs.

(10) Alice and Bob measure the EPR pairs in the
$ | 0\rangle$, $| 1 \rangle$ basis to obtain a shared secret key.

{\it Remark}: As discussed by Shor and Preskill, Alice should
also scramble the qubits by a random permutation before
sending them to Bob. Such a scrambling extends the tolerable
error rate from about $7$ percents set by Mayers \cite{mayersqkd} to
about $11$ percents in Shor-Preskill's proof.
We shall assume that this is done.

The above protocol consists of two steps: a) verification and b)
privacy amplification/error correction.
In step a), Alice and Bob verify by {\it random} sampling that
the error rate of the transmission is smaller than some prescribed
value. Otherwise, they abort. In step b), Alice and
Bob employ the property of CSS code to correct up to $t$ errors and
obtain privacy.

One can calculate the probability that the test on the check bits
is passed and yet the entanglement purification procedure on the
code bit fails. Since Eve does not know which qubits are used as
check bits and which as code bits, she cannot treat them
differently. In other words, the check bits provide a random
sample of all the bits. Moreover, since all relevant
measurements refer to the Bell-bases and thus commute with
each other, one can apply a {\it classical} random sampling argument
to estimate the number of errors. By choosing an appropriate
CSS code and $e_{max}$,
one can ensure that this probability is exponentially small in $n$.
The readers should refer to \cite{shorpre,squeeze} for details.

\subsection{Reduction to a quantum error-correcting code protocol}
Now, the above entanglement purification protocol only involves
one-way communication from Alice to Bob. It has been shown \cite{BDSW}
that any one-way purification protocol can be reduced to a quantum
error-correcting code protocol. i.e., Instead of Alice preparing
EPR pairs and sending halves to Bob, Alice prepares an encoded
quantum state with a quantum error correcting code and sends it to
Bob.

More concretely, suppose Alice and Bob start with $n$ perfect EPR pairs.
Suppose in step (9) Alice measures the eigenvalues of
$\sigma_z^{[r]}$ for each row
$r \in H_1$ and  $\sigma_x^{[r]}$ for each row
$r \in H_2$ and obtains the results, $x$ and $z$ respectively.
Her measurement will project the state of Bob into the CSS codespace
$Q_{x,z}$, which has basis vectors indexed by the coset of
$C_2$ in $C_1$. For $v \in C_1$, the corresponding codeword is
given by
\begin{equation}
v \to { 1 \over  | C_2|^{1/2}} \sum_{w \in C_2} ( -1)^{z \cdot w}
| x + v + w \rangle .
\label{eq:v2}
\end{equation}

The index, $x$, and $z$, defines a family of CSS
codes, $Q_{x,z}$, with equivalent error correcting capability.
In other words, each of them can correct up to $t$ phase errors
and $t$ bit-flip errors.

Also, Alice may measure her half of the EPR pair
before or after transmission. If she measures first, it will be the same
as she has chosen a random raw key
$k$
and encoded it by $Q_{x,z}$
(i.e., take $v = k$ in Eq. (\ref{eq:v2})).

\subsection{Reduction to BB84}
The property of CSS codes is used in
the reduction from a quantum
error correcting code protocol to BB84. Recall that the
bit-flip and phase error correction procedures decouple in a CSS code.
What if Alice and Bob simply drop the phase error correction procedure?
The resulting protocol is essentially BB84!

More concretely, since Bob does not really need the phase error
syndrome $z$ to extract the value of the shared key, there is no
reason for Alice to send it. Let us now consider the case when
Alice has obtained a value $k$ for the raw
key and
does not send $z$. We can take the average density
matrix of Bob, over all values of $z$, thus obtaining:
\begin{eqnarray}
~&{ 1 \over 2^n | C_2 | }  \sum_z \sum_{w_1, w_2 \in C_2}
(-1)^{(w_1 + w_2 ) \cdot z} \nonumber \\
~& \times | k + w_1 + x \rangle
\langle k + w_2 +x |  \nonumber \\
=& {{ 1 \over  | C_2 | }} \sum_{w \in C_2} | k + w + x \rangle
\langle k + w +x |.
\end{eqnarray}
This gives rise to a classical mixture of the states, $| k + w + x \rangle$
with $w$ randomly chosen from $C_2$.
Mathematically, the key extraction procedure is the same as the following
{\it classical} error correction/privacy amplification procedure:
Alice sends a random string $v$ to Bob and later
broadcasts $u+ v$ where $u $ is a random string in $C_1$.
The key is then the coset, $u +  C_2$, of $C_2$ in $C_1$.
Bob receives a corrupted string $v + e$.
He then substracts Alice's broadcast string $u+v$ from his
string to obtain $u+ e$. He corrects errors to find $u$ in $C_1$.
He then finds the final key to be $u + C_2$, which is a coset of $C_2$
in $C_1$.

\section{BB84 vs six-state scheme}
Let us look at Shor-Preskill's proof of security of BB84 more
closely by re-examining their underlying entanglement purification
protocol (EPP), Protocol~1. Recall from subsection
\ref{ss:classical} that one can employ the commuting observable
idea and only be concerned with probabilities of their
simultaneous eigenstates. In such a description,
one only considers the {\it diagonal} entries of the
density matrix with respect to the Bell-basis.
Furthermore, in the large $N$ limit (where $N$ is the
number of pairs of qubits), by random
sampling, one should only be concerned with the {\it average}
density matrix. Therefore, one can reduce the
whole problem of purification of a general $N$-pair
state in QKD to
the problem of purification of an ensemble of $N$ {\it identical
Bell-diagonal} states. In what follows, we will see that
the four entries in the density matrix have interpretations
in terms of the probabilities of a) no error, b) a bit-flip error,
but no phase error, c) a phase error, but no bit-flip error
and d) both bit-flip and phase errors. A natural
question to ask is: What are the correlations between the bit-flip
and phase errors?

We will now show that in Protocol~1, the bit-flip and phase errors
can be totally uncorrelated.
In the language of commuting observables and in the limit of
large number of pairs, let
us denote the effective density matrix by:

\begin{equation}
diag ( a, b, c, d).
\label{e:abcd}
\end{equation}

Here, we use the Bell-basis as in the notations of \cite{BDSW}.
(See also Eqs. (\ref{e:psi}) and (\ref{e:phi}).)
Now, the action of the Hadamard transform in Eq. (\ref{e:hadamard})
will permute the four matrix elements of Eq. (\ref{e:abcd}) into:

\begin{equation}
diag ( a, c, b, d).
\end{equation}

In Step~2 of Protocol~1, one applies randomly either the identity or
the Hadamard.
Averaging over the two cases: a) Identity and b) Hadamard,
we find that the effective average density matrix shared
by Alice and Bob after Step is of the form:

\begin{equation}
diag ( a, (b +c) /2, (b +c) /2, d)= diag (a, e, e, d),
\label{e:symmetric}
\end{equation}
where we define $e = (b +c) /2$.

As remarked earlier, the four entries represent, for each shared
pair between Alice and Bob, the four physical
possibilities respectively: a) No error;
b) bit-flip error, but no phase error; c) phase error, but no
bit-flip error; and d) both bit-flip error and phase error.
See, for example, \cite{BDSW} for details.
In the random sampling procedure---Steps~(5)-(8),
the sample bit error rate found by Alice and Bob will be
approximately $e+ d$. \footnote{We remark that, owing to
the symmetry between the two bases, only
a single bit error rate of the sample is required to
establish the security of the Shor-Preskill's procedure.
In other words, there is no need to employ a refined
data analysis studied in \cite{eff}.}
This leaves $d$ unconstrained
in Shor-Preskill's
proof of security of BB84 scheme. The implication is that
Alice and Bob cannot possibly know of
the correlations between the bit-flip and the phase error.
This is a serious limitation of the BB84 scheme.
In the worst case situation, the bit-flip and phase error
are independent. This corresponds to the
values, $e= (b +c) /2 = p (1-p)$ and $d = p^2$ for some
$ 0 < p < 1$.\footnote{Owing to symmetrization by the
Hadamard transform, two of the diagonal entries in
Eq. (\ref{e:symmetric}) are the same. This means that
the probability of having a bit-flip error but no
phase error is the same as that of having a phase error
but no bit-flip error. Now, suppose the two types of errors
are independent. They must occur independently with the
same probability $p$. This means that $e = p( 1-p)$ and
that $d =p^2$, as stated in the main text.}

Let us now consider the six-state scheme.
We will now show that the situation there is completely different.
Indeed, we will establish that, for the six-state scheme, the density matrix
is that given by a depolarizing channel and as such {\it does} have
correlations between bit-flip and phase errors.
It is this correlations between bit-flip and phase errors
that will give the six-state scheme an advantage over BB84.

As an analog of the Hadamard transform, which symmetrizes
between the two bases---$X$ and $Z$---in BB84, in the
six-state scheme we look for a symmetry operator that
will symmetrize between the three bases---$X$, $Y$ and $Z$.
We find the operator (see, e.g., Eq. (15) of \cite{gottesman}.)
\begin{equation}
T  = { 1 \over \sqrt{2} } \left( \begin{array}{cc} 1 & -i \\
 1 & i \end{array} \right),
\end{equation}
which cyclically permutes the three bases. i.e.,
\begin{equation}
T: X \to Y \to Z \to X .
\end{equation}

Suppose we apply either i) the identity operator; or ii) $T$; or iii) $T^2$
with equal probability to the density matrix
shown in Eq.~\ref{e:abcd}.
The average density matrix becomes
\begin{equation}
diag ( a, (b +c+ d ) /3, (b +c+ d ) /3, (b +c+ d ) /3) ,
\end{equation}
which is totally symmetric with respect to the three bases,
$X$, $Y$ and $Z$. This shows that the channel is
effectively a depolarizing channel \cite{BDSW}. More importantly,
the above four entries again represent the four possibilities:
a) No error;
b) bit-flip error, but no phase error; c) phase error, but no
bit-flip error; and d) both bit-flip error and phase error.
This implies that there are non-trivial correlations between
bit-flip and phase errors.

Such non-trivial correlations can be exploited to design
a six-state error correction/privacy amplification
protocol that tolerates a higher error rate
than Shor-Preskill's protocol for BB84. The key point is that,
the bit-flip error pattern and the phase error pattern are
no longer independent in the six-state scheme.
Therefore, given the same bit
error rate, the actual entropy of
the density matrix is smaller in the case of the six-state
scheme, as compared to the worst case situation in BB84.

More concretely, for the $n$ imperfect EPR pairs
shared by Alice and Bob, let us denote by the variable $\cal X$, the
phase error pattern and by $\cal Z$, the bit-flip error pattern.
Now, the entropy of the whole error pattern is given by
\begin{equation}
H ({\cal X}, {\cal Z}) = H({\cal X} ) + H({\cal Z}) - I ({\cal X} ; {\cal Z}).
\end{equation}
The fact that the phase and bit-flip error patterns are correlated
means that $I({\cal X};{\cal Z}) > 0$. Consider now the following
strategy of quantum error correction.

{\bf Subrountine~A: Modified ``random'' hashing procedure
with CSS codes}

(I) Alice and Bob apply a random hashing code on the ${\cal Z}$ variable only
to
identify the bit-flip error pattern.
Note that (slightly more than) $H(Z)$ rounds of random hashing is needed.

(II) Alice and Bob use the information on the bit-flip error pattern to
reduce their ignorance on the phase error pattern from $H({\cal X}) $ to
$H({\cal X}|{\cal Z}) = H({\cal X}) - I({\cal X};{\cal Z})=
H({\cal X},{\cal Z}) - H({\cal Z})$.

(III) Alice and Bob apply a random hashing code on the ${\cal X}$
variable only to identify the phase error pattern.
Note that only (slightly more than)
$H({\cal X},{\cal Z}) - H({\cal Z})$
rounds of
random hashing is needed.

{\it Remark}: Bennett, DiVincenzo, Smolin and Wootters (BDSW) \cite{BDSW}
have studied a random hashing scheme
in entanglement purification. Our random hashing code is
analogous, except that we restrict our attention to CSS codes.
Therefore, in (I), all the operators are chosen to
be tensor products of $Z$ operators only and in (III), $X$ operators only.
Nonetheless, in the asymptotic limit of large number of pairs,
our scheme is equally efficient as the original random hashing scheme
by BDSW. In BDSW, it was shown that the scheme gives non-zero
rate of distilled entanglement when the fidelity $f > 0.81071$.
Since $f = 1 - 3p/2$ for a depolarizing channel, this corresponds to
a bit error rate of about 12.6 percents.

{\it Remark}: By adopting a modified version of the
DiVincenzo-Shor-Smolin code \cite{DSS}, a slightly
higher error rate of about 12.7 percents can be tolerated in
the six-state scheme. DSS code consists of the concatenated of
a non-random (cat) code with random hashing code. It is
one of the few examples of a so-called degenerate code and gives
better performance than any known non-degenerate code.
Note that the non-random (cat) code is a CSS-code. Since we have
given a modified version of the random hashing code that
is a CSS code, the concatenated code can, therefore, also be
modified into a CSS code. While the actual improvement---12.7 percents
vs 12.6 percents---is quite small, the result is
conceptually interesting because it shows that a degenerate
code can be employed in the underlying entanglement purification
protocol in i) establishing a secure error-correction/privacy
amplification protocol for the six-state QKD scheme and
ii) tolerating a higher error rate than any known non-degenerate
codes.

\section{Protocol for secure six-state QKD scheme}
Following our discussion in the last section, we
now give the details of our procotol for secure
six-state QKD scheme. We claim the following modified
QKD scheme
is secure. For conciseness, we omit the steps that
are identical to Protocol~1. We replace some of the steps of Protocol~1
by the following. 

{\bf Protocol~1'}: {\bf QKD based on entanglement purification}

(0') Alice and Bob decide on a large positive integer $n$ and a maximal
number $e_{max}$ of check bit errors that they tolerate
in the protocol.

(2') Alice selects a random $2n$-trit $t$, and performs
$I$, $T$ or $T^2$ on the second half of each EPR pair
if (the component of) $t$ is $0$, $1$ or $2$ respectively.

(6') Alice announces the trit string $t$ and which $n$ EPR pairs
to be check bits.

(7') Bob performs $I$, $T^{-1}$ or $T^{-2}$ on the qubits depending
on the value of (the component of) $t$.

(9') Alice and Bob apply Subrountine~A (the modified random
hashing procedure
with CSS codes discussed in the last section) to correct the
(correlated) bit-flip and phase errors. They broadcast their results.
Bob transforms his state accordingly to obtain $m$ nearly perfect
EPR pairs (which are shared with Alice).

The proof of security is analogus to Shor-Preskill's proof.

\subsection{Reduction to six-state protocol}
Furthermore, the various reduction arguments of Shor-Preskill directly
carry over and reduce Protocol~1' to
a six-state protocol. From the security of Protocol~1', we have proven
the security of the six-state protocol. 

More specifically, the quantum key distribution Protocol~1', which
is based on a one-way entanglement
purification protocol, is mathematically equivalent to a protocol
based on a class of CSS code. Furthermore, by the virtue of CSS codes,
the phase error-correcting procedure is essentially
decoupled from the bit-flip error-correcting procedure. 
Since the phase errors do not affect the
value of the final key, the phase error correction procedure can be
simply dropped. Put in another way, Alice could have done the
procedure with {\it any} CSS code in the same family (they are all
related by phase errors to one another). Mathematically, the
mixture of the CSS codes in the family is equivalent to a classical
code (with the corresponding error correction and privacy amplification
procedure). Therefore, the protocol can be reduced to the a simple
``prepare and measure'' protocol, namely the six-state scheme.
The maximal tolerable bit error rate of the six-state scheme
with our error-correction/privacy amplification procedure is
12.6 percents for modified random hashing with CSS codes
(and 12.7 percents if we employ a modified DSS code described
in the last Section).

Inamori \cite{inamori6}
has recently proposed a proof based on a different approach
which gives a higher tolerable
error rate of about $13\%$. However, unlike the present
proof, Inamori's proof
requires two-way communications between Alice and Bob.

In conclusion, we have proven the security of the six-state
quantum key distribution up to a bit error rate of 12.7 percents.

\section{Concluding remarks}
We shall conclude with a few remarks.

\subsection{Efficient six-state scheme and proof of its unconditional
security}
In the six-state scheme, Alice and Bob independently and randomly
choose between three bases. Therefore, two-thirds of the times they
disagree and have to throw away their polarization data. We remark
that they can improve the efficiency of scheme substantially by
choosing the three bases with different probabilities, say
$\epsilon, \epsilon$ and
$ 1 - 2 \epsilon$. This ensures that the efficiency is
greater than $ ( 1 - 2 \epsilon)^2$. As $\epsilon \to 0$, the efficiency
asymptotically
goes to $100\%$.

Whereas in the standard six-state scheme
the computation of only a single error rate is
required for its proof of security,
for this efficient scheme to be secure,
it is now necessary to use a refined data analysis \cite{eff}.
One should divide up the data according to the various bases in
which they are transmitted and received and compute the error rate for
each basis {\it separately} and demand that all the error rates are small.

Note that the scheme is insecure when $\epsilon$ is exactly zero.
The constraint on $\epsilon$ has been discussed.
Basically, it is necessary that
$N \epsilon^2 > m$ where $N$ is the total number of
photons transmitted from Alice to Bob and $m$ is the minimal number of
photons needed for an accurate estimation of the error rate of the data.
The numerical value of $m$ must scale at least
as $\log N$. But, it is a priori unnecessary for it to scale
linearly with $N$.
We remark that the unconditional
security of the efficient four-state scheme
has been proven in \cite{eff}. It is straightforward to apply the
techniques developed there to prove the unconditional security of the
efficient six-state scheme up to the same error rate of 12.7 percents.

\subsection{Security of other QKD schemes}
The security of some other QKD schemes remains to be
explored. In particular, it would be interesting to
study the security of the B92 scheme \cite{B92} with noises.
It is not entirely obvious to us how the Shor-Preskill's techniques
can be applied to B92.

\subsection{Real life issues}
\label{ss:real}
Our result only applies to an idealized situation. In a real experiment,
the source of EPR pairs are imperfect; the channel is lossy and
the detector efficiency is far from perfect. It would be interesting to
explore the security of the six-state in a real world situation.
For BB84,
some works along those lines have been done by
researchers including L\"{u}tkenhaus \cite{lut,gilbert}.

In a recent preprint, Inamori, L\"{u}tkenhaus and Mayers \cite{coherent}
have proposed a proof
of security of a weak coherent state implementation of the BB84 scheme.
A key assumption is that, given any quantum signal, independent of the
basis of measurement chosen by Bob, Bob's detection efficiency
stays the same. (One can imagine that Bob chooses his measurement
basis by pushing a button. Then, independent of which
button his pushes, it is assumed that the measurement
will with the same probability be successful. In other words,
the signals {\it cannot} behave differently according to the
basis chosen by Bob. Cf. Trojan Horse attack in the next subsection.)
This assumption is closely related to the detectors' loophole problem
in the testing of Bell's inequalities. Given rather imperfect detectors,
testing of Bell's inequalities often assumes that the detected
sample provides a fair representation of all the signals, detected or not.

The Shor-Preskill's proof is a fine theoretical result. However,
if one would like
to apply the result, one needs to make sure that the
amount of computing power required is reasonable.
It is not entirely clear to us that this is the case.
Some discussion
has been made in \cite{gilbert}.

\subsection{Trojan Horse problem}
In proofs of security of QKD schemes, it is often
assumed that the signals transmitted from Alice to Bob lives
in a two-dimensional space. How can one be sure that there is
no hidden Trojan Horse in the signal? For instance, the
signal may, in principle, be made up of two parts, one is
the usual quantum signal, the other is a robot that will
explore Alice or Bob's system and tell the first part of the
signal to behave differently according to, for example, the basis of
measurement actually employed by Bob.
Notice that the Trojan Horse can break the quantum crypto-system
without directly leaking out information from Bob's laboratory to Eve!

One might naively think that QKD provides more room for
the Trojan Horse attack. Fortunately, it
has been pointed out (Note~21 of \cite{qkd}) that this Trojan
Horse problem in quantum cryptography is no worse than
in classical cryptography:
By using teleportation, any
quantum signal can be reduced to classical one. Therefore,
Alice and Bob only need to receive classical signals anyway.
This teleportation trick
requires only the experimental implementation
of teleportation, rather than a full-blown quantum computer.

\subsection{Bell's inequality with untrusted imperfect apparatus}
Another question is whether Alice and Bob can buy their
quantum cryptographic devices
from untrusted vendors and verify their
security by doing some simple testing themselves. By assuming that
Alice and Bob's laboratory can be sufficiently shielded
from the environment, a procedure to prove security based only
on input/output probabilities (that corresponds to a choice of
several local measurements by Alice and Bob and the corresponding
measurement outcomes) has been provided for the case of perfect
EPR pairs \cite{mayersyao}. It would, thus, be interesting to
generalize the result to the case of imperfect EPR sources and
measuring apparatus. (See also
Subsection~\ref{ss:real}.)
This line of research can also be re-phrased
as a generalization of Bell's inequality to the case of a limited
amount of entanglement. The question there becomes: given a fixed
amount of entanglement, how far can Bell's inequality be violated?
Conversely, given some experimental violations of Bell's inequality
on some random sample, can one deduce the minimal
amount of entanglement shared by Alice and Bob?
Interesting questions include what type of privacy amplication/error
correction procedures
can be employed to prove unconditional security
in this untrusted situation.

In summary, there is no doubt in
our mind that QKD provides a fertile real-life playground for
the various concepts in quantum information theory.
Moreover, these interactions between theory and practice will
most likely inspire new research avenues on both sides.

{\it Notes Added}: Recently, Gottesman and Lo \cite{two} have proven the
security of the six-state scheme up to a bit error rate of about
23 percents. However, their method employs two-way classical
communications between Alice and Bob. They also show that,
by allowing two-way classical communications, BB84 scheme can be
made unconditionally secure up to an error rate of 17 percents.

\section{Acknowledgments}

We thank helpful discussions with various colleagues including
Charles Bennett, David DiVincenzo,
Nicolas Gisin, Daniel Gottesman, Hitoshi Inamori,
John Preskill, Peter Shor, John Smolin and Barbara
Terhal. We particularly thank Debbie Leung for
pointing out some minor errors in the earlier version of this
paper, Gerard Gilbert for bringing Ref. \cite{gilbert} to
our attention and Norbert L\"{u}tkenhaus for helpful
suggestions and critical comments.

\end{document}